# High aspect ratio metal microcasting by hot embossing for X-ray optics fabrication

L. Romano[a,b,c], J. Vila-Comamala[b,c], M. Kagias[b,c], K. Vogelsang[b], H. Schift[b], M. Stampanoni[b,c], K. Jefimovs[b,c]

[a] Department of Physics and CNR-IMM- University of Catania, 64 via S. Sofia, Catania, Italy;
[b] Paul Scherrer Institut, 5232 Villigen PSI, Switzerland;
[c] Institute for Biomedical Engineering, University and ETH Zürich, 8092 Zürich, Switzerland

**Abstract**

Metal microstructured optical elements for grating-based X-ray phase-contrast interferometry were fabricated by using an innovative approach of microcasting: hot embossing technology with low melting temperature (280°C) metal alloy foils and silicon etched templates. A gold-tin alloy (80w%Au / 20w%Sn) was used to cast micro-gratings with pitch sizes in the range of 2 to 20 µm and depth of the structures up to 80 µm. The metal filling of the silicon template strongly depends on the wetting properties of the liquid metal on the groove surface. A thin metal wetting layer (20 nm of Ir or Au) was deposited before the casting in order to turn the template surface into hydrophilic with respect of the melted metal alloy. Temperature and pressure of the hot embossing process were optimized for a complete filling of the cavities in a low viscosity regime of the liquid metal, and for minimizing the shear force that might damage the silicon structures for small pitch grating. The new method has relevant advantages, such as being a low cost technique, fast and easily scalable to large area fabrication.

1.  **Introduction**

Metal microstructures are nowadays of great interest for sensors and electromechanical devices [1], manufacturing molds for nanoimprint lithography (NIL) [2,3], and for many optics applications, such as plasmonic devices, photonic crystals [4] and X-ray optics [5]. In particular, interferometric X-ray phase contrast imaging based on diffraction gratings requires the fabrication of high aspect ratio (HAR) metal microstructures [6]. Grating-based X-ray phase-contrast Interferometry has shown to provide a much higher image contrast than conventional absorption-based X-ray radiography imaging [5]. This has a high application impact in material science and medicine for imaging of weakly absorbing (low Z) materials and soft tissues [5]. The essential part of the traditional interferometer consists of one phase and two absorption gratings. The main challenge is the fabrication of the absorption gratings, whose quality and aspect ratio (AR) strongly affect the quality of the generated images. There is the need to fabricate gratings with i) high aspect ratio (in the range of 100:1, structural width in the micrometer range); ii) large area (mammography, e.g., asks for a field of view of 200x200 $mm^2$) and iii) good uniformity (no distortions and change in the period and height over the whole grating area). Absorption gratings are usually fabricated by metal electroplating (typically of gold, which is one of the highly absorbing material for X-rays), into high aspect ratio grating templates produced by LIGA [7] or deep silicon (Si) etching [8].

These existing processes and materials lead to high fabrication cost and low yield, limiting the broad utilization of X-ray grating interferometry for industrial and medical applications. Therefore, the mass production of HAR and large area absorption gratings with metal microstructures through a simple process is essential for commercialization of GI based x-ray imaging systems.

Metal microstructures are typically manufactured using forging [9], electroplating [8], micro powder injection molding [1] or casting [2,10]. Forging can produce small-scale structures in ductile metals [4]; the depth is usually limited to the micrometer range [4,9]. Microscale metal electroplating is less expensive than forging, but is usually slower. Micro powder injection molding is faster than electroplating, but it is currently limited to 20 μm structure width sizes. Metal nanostructures have been produced in thin metal film by using NIL [4,11,12]. While metal microstructures can be manufactured using casting, the direct imprint in metal has not been deeply investigated in comparison to the other approaches. Microcasting for X-ray absorption gratings has been developed by using molten bismuth (melting temperature 271 °C) via capillary action and surface tension [13,14]. However, the low density (9.78 $g/cm^3$, atomic number 83) of bismuth requires much higher (factor of 1.7 at 30 keV) AR structures to get an absorption level comparable to that of gold (density 19.32 $g/cm^3$, atomic number 79). Gold microcasting can be very expensive and not easy to scale up on large area, not only for the required bulk quantity of liquid metal but also for the high melting temperature of gold (1064 °C). There are some other metal alloys which have the benefits of low temperature liquid phase and high X-ray absorption. For example, several tin alloys have been developed for metal bonding technology [15,16], like gold-tin and lead-tin alloys. These materials are commercially available in a wide range of foil thicknesses from few tens to hundreds of micrometers.

In this paper, we propose the use of hot embossing with eutectic gold-tin (Au 80w% / Sn 20w%) foils into Si templates. The low melting point at 280 °C, its relatively high density 14.7 $g/cm^3$, make Au-Sn alloy an attractive material for casting X-ray absorption gratings. For example, it requires only a factor of 1.23 higher structures than pure gold at 30 keV to get the same absorption. Moreover, the hot embossing technique is a well-known technique for large area silicon processing being largely used as a tool for NIL technology [17]. In this publication, due to narrow process window for the phase transition between solid and liquid phase, we characterize the physical process as microcasting but the technical method as hot embossing.

## 2. Material and methods

Figure 1.a shows the schematic process flow for preparing the metal microstructures into Si templates. First, a pattern was realized by conventional UV photolithography in a positive photoresist (MicroChem S1805) Si gratings with duty cycle 0.5, pitch in the range of 2 – 20 µm, and depth in the range of 25 – 80 µm were fabricated by deep reactive ion etching [18] with the so-called Bosch process [19] or by Metal Assisted Chemical Etching (MACE) [20,21]. The 100 mm diameter Si wafers were <100> n-type (0.005-0.01 Ωcm) and p-type (1-30 Ωcm) for Bosch etch and MACE, respectively. An example of Si grating produced by the Bosch process is reported in Figure 1.b, the grating was cleaved in order to obtain a cross section image by Scanning Electron Microscopy (SEM). Seedless electroplating [22] was used to grow a conformal thin layer of Au on the Bosch etched grating (see Figure 1.c) to improve the wettability of the liquid metal on the template during the casting process. Due to the high resistivity of the wafer, the seedless electroplating is not possible on MACE gratings, so Atomic Layer Deposition (ALD) was used to realize a conformal coating of 20 nm of iridium (Ir). Afterwards square gratings of 20x20 mm$^2$ up to 70x70 mm$^2$ were cut out from the wafer. The metal casting was performed in a Jenoptik HEX 03 hot embossing tool, in vacuum (pressure in the range 100-500 Pa), by pressing a metal foil with same size as the grating in contact with the grating surface at the melting temperature. The pressure was varied from 1 MPa to 12 MPa. The temperature was controlled with a precision of 2° C and the maximum temperature for this system is 320 °C, therefore metal foils of material with melting point lower than 320 °C can be used. We used a metal foil of eutectic Au-Sn (80w%Au / 20w%Sn with eutectic temperature of 280°C) alloy [23] from Ametek with thickness of 25 and 50 µm. The thickness of the Au-Sn alloy foil was chosen depending on the profile depth, matching the cavity volume of the grating and minimizing the excess of material. For example, for a duty cycle of 0.5, the thicknesses of the foil should typically be a half the depth of the grooves.

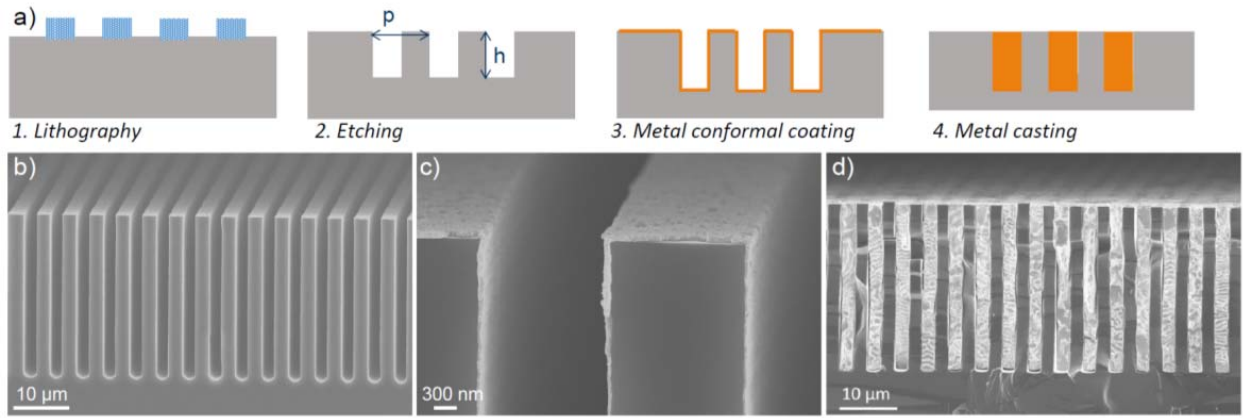

**Figure 1.** a) Schematic of the full fabrication process of gratings (not to scale). The following SEM cross section images show some examples of the reported process steps: b) SEM cross section detail of 70x70 mm$^2$, 40 µm deep Si grating etched by Bosch (representative of step 2); c) seedless Au electrodeposition (20 nm Au) to produce a wetting layer on the entire Si surface (representative of step 3); d) Au-Sn alloy casting in the Si template after hot embossing (representative of step 4) realized with applied force of 20 kN on 70x70 mm$^2$ grating.

A typical hot embossing experiment [24] is reported in the schematic of Figure 2.a (not in scale). A 1 mm thick sheet of silicone rubber (PDMS, i.e. polydimethylsiloxane) was used as a cushion layer for pressure equilibration, it is sufficient to smoothen out any kind of unevenness, e.g. caused by substrate bow and warp or even dust particles. In order to avoid sticking of the PDMS [25], a polyimide foil was used on both sides of the PDMS as an anti-sticking layer. A Si chip with thickness of 500 µm was used as a flat surface to apply the pressure on the metal foil, a polyimide foil between the metal and the Si chip helps to easily detach the casted grating at the end of the process. Figure 2.b reports the measured force and tool temperature as a function of time in a typical process of hot embossing on a grating with size of 20 x 20 mm$^2$: 1) after evacuating the embossing chamber (in the range 100-500 Pa), the hot plates were heated up from room temperature and 2) a touch force of 300 N was applied while the heating took place with about 15 °C/min; 3) a force of 5 kN was applied once the substrate reached the melting temperature of the metal foil; the applied force was maintained for few minutes and then 4) the system was cooled down to room temperature and finally, 5) the force was released.

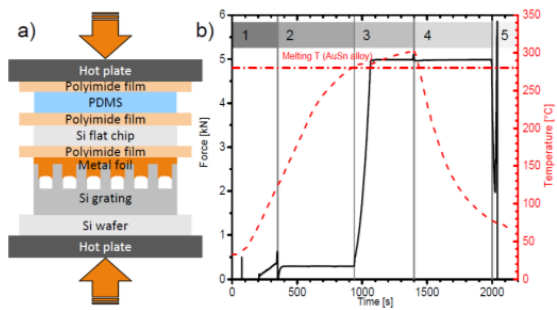

**Figure 2.** a) Schematic representation (not to scale) of the hot embossing configuration; b) *In-situ* monitored parameters during a typical hot embossing experiment as a function of time: the touch force is reported on the left scale by black continuous line; the temperature measured on the substrate is reported on the right scale and shown in dashed red line (the melting temperature of the Au-Sn alloy is also indicated as dash-dot line, 280°C). The various phases of the process are indicated: 1) the system is in vacuum and the hot plates are heating up; 2) the touch force of 300 N is applied; 3) the embossing force of 5 kN is applied once the melting temperature of the metal foil is reached and the system is kept compressed at a temperature slightly higher than the melting temperature; 4) the system is cooling down while the force is still applied; 5) the force is finally released.

3. Results & Discussion

In order to ensure that the metal does not flow sideways, the geometrical conditions need to be fulfilled, which means that the foil thickness has to match with the cavity volume of the grating. On the other hand, to achieve complete filling of the lines a wetting layer was used and the casting process was performed in vacuum. We applied different pressure and temperature ramping to optimize the casting process. The good control of pressure, temperature and speed of the system allowed to hot emboss at the melting point of the metal.

Figure 1.d shows the Si trenches filled with bubble-free gold-tin alloy in a 4.8 µm pitch grating with a size of 70x70 mm$^2$. The metal foil was 25 µm thick and the applied force during hot embossing was 20 kN (equivalent to a pressure of 8 MPa on the 70x70 mm$^2$ square and due to the 0.5 duty cycle of the grating). The hot embossed grating was inspected with optical microscopy (not shown), revealing a very uniform filling of the trenches over the full area. The quality of the metal grating was investigated by SEM. For the 4.8 µm pitch grating, the trenches were completely filled with no empty cavities. The excess material flowed sideways on the top of the grating leaving a clean top surface with a very thin residual layer of~ 100 nm.

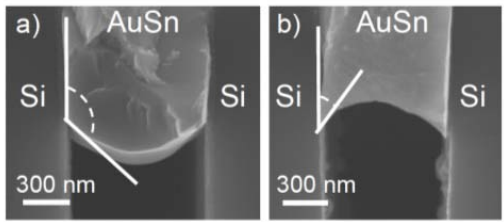

**Figure 3.** a) example of hydrophobic behavior of Au-Sn in Si grooves (wetting angle $\vartheta$~120°); b) example of hydrophilic behavior ($\vartheta$~45°) of Au-Sn in Si grooves coated by a metal wetting layer (in this case Au by electroplating). The wetting angle $\vartheta$ is sketched.

The wetting properties of the liquid metal both on top and on the trench wall surface were extremely important for uniformly filling the Si grooves. When the liquid metal did not uniformly wet the available Si surface, the metal was unequally distributed over the grooves filling only some of them. This is most probably due to the pressure applied and not by capillary forces, since the surfaces are non-wetting. The metal being pressed in some grooves was not reaching the bottom of the grating, and more metal could only be injected by bending the Si ridges towards the non-filled grooves. Since the grooves filled with metal were able to bend the silicon lines, the final Au-Sn ridges exhibit a larger size on the top of the grating with respect to the bottom. In some cases of a partial filling of the Si grooves, we observed the non-wetting behavior of the liquid metal. Figure 3.a shows the SEM image of a Si groove partially filled with Au-Sn metal. Even if the image is obtained after the solidification of the metal, it is possible to have an indication of the wetting angle $\vartheta$ during the liquid phase, being $\vartheta$>90° the liquid stopped flowing into the trench leading to a partial filling of the groove. The angle of $\vartheta$~120° is probably smaller than the actual wetting angle due to the thermal contraction of the Au-Sn alloy during cooling. On the contrary, if a preliminary conformal metal coating of the Si surface was realized by Au electrodeposition or Ir ALD, the liquid metal was drawn into the grooves by capillary forces during the hot embossing and it could uniformly wet the internal surface of the grooves, with a wetting angle of $\vartheta<90°$ (Figure 3.b). Also here the angle of $\vartheta\sim45°$ is only a rough indicator of the wetting angle in the liquid phase. Once the proper control of the wetting of the liquid metal by preliminary coating of the Si surface is achieved, high aspect ratio and very deep structures can be realized my casting with Au-Sn alloy. An example of grating with pitch size p = 6 µm and depth h = 80 µm is reported in Figure 4. In this case, due to the high value of depth (h = 80 µm) the wetting layer was realized by coating the Si grating with an Ir layer of 20 nm by ALD before the Au-Sn casting.

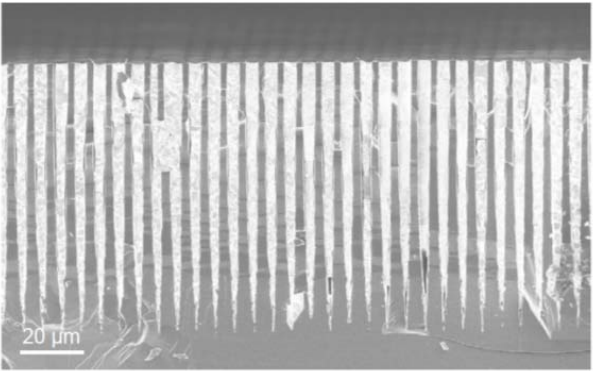

**Figure 4.** 70x70 mm$^2$ Bosch grating (p = 6 µm, h = 80 µm, aspect ratio 25:1) with 20 nm Ir wetting layer by ALD and Au-Sn casting obtained with applied force of 20 kN. The sharp tips at the bottom of the grooves are due to the silicon reactive ion etching.

The role of the hot embossing parameters, such as pressure, temperature and speed of the mechanical compression, are relevant for tuning the process on the desired grating size and aspect ratio and obtaining a good metal filling of the Si template. Since the Au-Sn foil is a eutectic alloy, the melting temperature is the lowest possible over all of the mixing ratios for the involved component species. The temperature control is essential to apply the pressure in the exact moment of solid-liquid transition of the metal foil. In the liquid phase the viscosity of the metal is minimal. Once solidified upon cooling in the hot embossing system, the metal does not usually melt again at the eutectic temperature of the original foil. This might be due to partial dissolution of the wetting layer of Au or Ir and changing composition of the alloy or precipitation of the elements of the alloy, which both lead to a rising of the melting temperature. Therefore, the external pressure has to be quickly applied at the moment of phase transition. The applied force has to ramp up to the final value very fast when the melting temperature is reached. It is also very important that the system is kept compressed during the entire process. Once the metal foil melts and the metal is able to flow, the press tries to maintain the applied force by moving the upper plate towards the lower one until the gap is closed. This is needed to complete the filling of the groves but also enables the liquid metal to flow sideways over the top surface of the template with a relevant loss of material. The hot embossing machine has the right set up to fulfill these requirements. The plot of Figure 2.b shows that the full embossing force is reached in about 100 s after the force ramping is started. The applied force ramp rate is limited partly by the compression of the PDMS stack and the squeezing of the molten alloy into the trenches of the grating. Although the stack is heated from both sides with the aim to achieve an isothermal process (with identical temperature at the bottom and the top of the stack), at the time of initiating the pressing the actual temperature at the interface between the Si grating and the metal foil is probably still slightly lower than the 280 °C measured by the sensors (about 10 mm within the metal plates of the press). Therefore the actual filling probably already takes part during the ramping. The metal alloy undergoes a rapid phase transition from solid to liquid, i.e. the filling is probably initiated while the metal is still in an inhomogeneous phase and then quickly – due to its high thermal conductivity – converts into the liquid phase. Once the material is entirely liquid, it will both be further squeezed into cavities and sideways. Therefore it is essential that most of the vertical filling of the groove cavities is finished before the excess material is removed by lateral flow. Since a homogeneous filling can be obtained from the center to the boundaries of a large 70x70 mm$^2$ grating, and taking into account the tolerances of both the thermal process and composition of the alloy, it seems that the process window for casting gratings in Au-Sn alloy is large and can probably be even extended. Although we do not know the exact viscosity of the AuSn alloy near its melting temperature (for temperatures clearly above we can assume a viscosity of around 1 mPa s [16,26]), such a large process window has been found similar to the low viscosity regime found in UV-NIL (5 mPa s), might also reach out to higher viscosity regimes in UV-NIL (50 mPa s), but probably not to the high viscosities used in thermal NIL ($10^3$-$10^7$ Pa s) [3,27]. It is also possible that – due to the direct heat transfer – the melting starts at the grating ridges in contact to the metal foil and that they indent into the foil which further inhibits the metal to flow sideways. Both a high uniformity of pressure and the wetting over the area of the grating is required to obtain a uniform metal casting. The maximum applied pressure has to be tuned in order to obtain a good filling of the metal into the template but also to minimize the shear force and the eventual damage of the high aspect ratio silicon structures. The latter is especially relevant as thinner structures are more prone to bending. Therefore for smaller grating pitch sizes high initial pressures and any shear forces due to flow or mechanical misalignment have to be avoided. Therefore, the embossing forces were reduced down to 500 N (which results in a pressure of ~1 MPa for 20x20 mm$^2$ gratings).

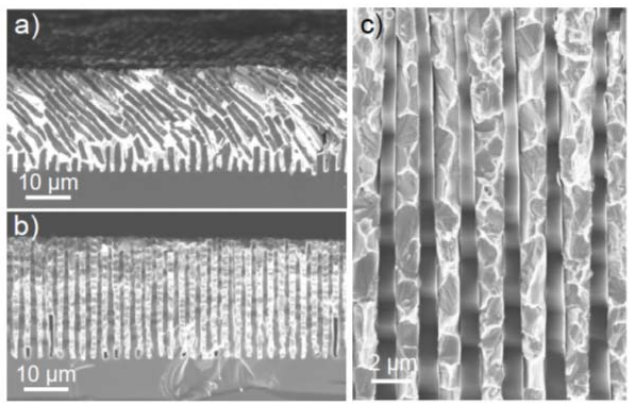

**Figure 5.** 20x20 mm$^2$ MACE grating (p = 2.4 µm, h = 30 µm, aspect ratio 25:1) with 20 nm Ir wetting layer by ALD and Au-Sn casting obtained with applied force of 5 kN (a) and 500 N (b). 20x20 mm$^2$ Bosch grating (p = 2 µm, h = 40 µm, aspect ratio 40:1) with 20 nm Au wetting layer by electroplating and Au-Sn casting obtained with applied force of 500 N.

Some examples of the role of the pressure are reported in Figure 5.a and 5.b. A grating produced by MACE with pitch size p=2.4 µm and aspect ratio of 25:1 was hot embossed with an applied force of 5 kN (Fig. 5.a) resulting into a good metal filling of the template, but Si lines were broken due to the excess of shear force. Figure 5.b shows that it is possible to obtain a good metal filling with preserving the Si structures on the same grating by reducing the applied force of one order of magnitude (500 N). With the same approach an aspect ratio of 40:1 was achieved for a 2 um pitch grating (Figure 5.c shows a detail in high magnification of the grating cross section). Some voids at the bottom of the grooves (Fig. 5.b) can be observed, this is probably due to an imperfect coating of the Ir wetting layer in some regions of the bottom surface, indicating once more the relevance of the wetting layer for the uniformity of the template filling.

In the experimental range of pitch size (2 – 20 µm) and depth (25 – 80 µm) we did not observe any structural limit of the technique, once the metal is liquid and wets the trench surface, temperature and pressure can be set in order that the metal uniformly fills the silicon template. Since the wetting layer plays a critical role, we can speculate that indeed this could represent the limiting factor for scaling up the aspect ratio. ALD conformality is reported to perform the aspect ratio of 1000:1 [28], further experiments are needed to extend the range of aspect ratio for the metal casting in order to access the structural limit of the technique.

First experiments indicated that the Au-Sn casted gratings perform well in X-ray grating interferometry. Whether they are comparable to that of Au electroplated gratings will be subject of further investigations. These preliminary results demonstrated that the microcasting is a viable technology for the fabrication of metal HAR microstructures for X-ray absorption gratings application, and it is promising to be applied to other metal alloys. In particular, Pb alloys are promising, exhibiting low cost, low temperature melting point (328 °C for pure Pb), and high density (11.34 g/cm$^3$ for pure Pb). Pb requires a factor of 1.50 higher structures than gold at 30 keV to get the same X-ray absorption. Lead alloys (92.5w% Pb - 5w% Sn - 2.5w% Ag and 95w% Pb - 5w% In, melting point at 290 °C and 314 °C, respectively) have been tested on 4.8 um pitch gratings (not shown) with good quality filling. Complete filling was obtained for the range of pressures 1-12 MPa, with lower pressures more appropriate to avoid damage for delicate structures.

4. Conclusions

Microstructured metal gratings with pitch size in the range of 2-20 µm were realized with high aspect ratio (up to 40:1) by using a microcasting technique with Au-Sn alloy (80w%Au / 20w%Sn) foils into Si templates with a typical hot embossing process time of approximately 30 min. A metal conformal wetting layer was necessary to obtain a good filling of the Si trenches with the molten Au-Sn alloy. Hot embossing parameters (sample set up, vacuum, applied force and temperature ramps) were optimized for minimizing the damage of the structure and for enabling a complete uniform filling. We demonstrated the fabrication of 70×70 mm$^2$ Au-Sn gratings on Si 100 mm diameter wafers. Similar process could be used for microcasting of other metals and alloys with low melting temperature. The reported results open the way to a new rapid fabrication route for large area metal grating with micrometer periods. Such gratings are key components for X-ray phase sensitive interferometric systems. By achieving a low cost and high yield fabrication process the applicability of X-ray grating interferometry in the medical and industrial field will be boosted. We believe that the method can also provide alternative solutions for rapid manufacturing of MEMS and microcomponents, such as the watch industry.

**Acknowledgements**

This work has been partially funded by the ERC-2012-STG 310005-PhaseX grant, ERC-PoC-2016 727246-MAGIC grant and the Fondazione Araldi Guinetti. We would like to thank for their valuable collaboration and contributions: S. Stutz, V. Guzenko and C. David from PSI-LMN, C. Arboleda and Z. Wang from PSI-TOMCAT.

**References**

[1]     G. Fu, S. Tor, N. Loh, B. Tay & D. E. Hardt, A micro powder injection molding apparatus for high aspect ratio metal micro-structure production, Journal of Micromechanics and Microengineering 17 (2007) 1803.
[2]     A. H. Cannon & W. P. King, Casting metal microstructures from a flexible and reusable mold, Journal of Micromechanics and Microengineering 19 (2009) 095016.
[3]     H. Schift, Nanoimprint lithography: An old story in modern times? A review, Journal of Vacuum Science & Technology B 26 (2008) 458-480.
[4]     S. Buzzi, F. Robin, V. Callegari & J. F. Löffler, Metal direct nanoimprinting for photonics, Microelectronic Engineering 85 (2008) 419-424.
[5]     T. Weitkamp, A. Diaz, C. David, F. Pfeiffer, M. Stampanoni, P. Cloetens & E. Ziegler, X-ray phase imaging with a grating interferometer, Opt. Express 13 (2005) 6296-6304.
[6]     T. Thüring, M. Abis, Z. Wang, C. David & M. Stampanoni, X-ray phase-contrast imaging at 100 keV on a conventional source, Scientific Reports 4 (2014) 5198.
[7]     J. Mohr, T. Grund, D. Kunka, J. Kenntner, J. Leuthold, J. Meiser, J. Schulz & M. Walter, High aspect ratio gratings for X-ray phase contrast imaging, AIP Conference Proceedings 1466 (2012) 41-50.
[8]     C. David, J. Bruder, T. Rohbeck, C. Grünzweig, C. Kottler, A. Diaz, O. Bunk & F. Pfeiffer, Fabrication of diffraction gratings for hard X-ray phase contrast imaging, Microelectronic Engineering 84 (2007) 1172-1177.
[9]     Y. Chen, Y. Zhou, G. Pan, E. Huq, B.-R. Lu, S.-Q. Xie, J. Wan, Z. Shu, X.-P. Qu, R. Liu, S. Banu, S. Birtwell & L. Jiang, Nanofabrication of SiC templates for direct hot embossing for metallic photonic structures and meta materials, Microelectronic Engineering 85 (2008) 1147-1151.
[10]    G. Baumeister, B. Okolo & J. Rögner, Microcasting of Al bronze: influence of casting parameters on the microstructure and the mechanical properties, Microsystem Technologies 14 (2008) 1647-1655.
[11]    H. L. Chen, S. Y. Chuang, H. C. Cheng, C. H. Lin & T. C. Chu, Directly patterning metal films by nanoimprint lithography with low-temperature and low-pressure, Microelectronic Engineering 83 (2006) 893-896.
[12]    S. W. Pang, T. Tamamura, M. Nakao, A. Ozawa & H. Masuda, Direct nano-printing on Al substrate using a SiC mold, Journal of Vacuum Science & Technology B 16 (1998) 1145-1149.
[13]    Y. Lei, L. Xin, L. Ji, G. Jinchuan & N. Hanben, Improvement of filling bismuth for x-ray absorption gratings through the enhancement of wettability, Journal of Micromechanics and Microengineering 26 (2016) 065011.

[14] Y. Lei, D. Yang, L. Ji, Z. Zhigang, L. Xin, G. Jinchuan & N. Hanben, Fabrication of x-ray absorption gratings via micro-casting for grating-based phase contrast imaging, Journal of Micromechanics and Microengineering 24 (2014) 015007.

[15] M. A. Schmidt, Wafer-to-wafer bonding for microstructure formation, Proceedings of the IEEE 86 (1998) 1575-1585.

[16] G. S. Matijasevic, C. C. Lee & C. Y. Wang, AuSn alloy phase diagram and properties related to its use as a bonding medium, Thin Solid Films 223 (1993) 276-287.

[17] H. Schift, C. David, M. Gabriel, J. Gobrecht, L. J. Heyderman, W. Kaiser, S. Köppel & L. Scandella, Nanoreplication in polymers using hot embossing and injection molding, Microelectronic Engineering 53 (2000) 171-174.

[18] S. Rutishauser, M. Bednarzik, I. Zanette, T. Weitkamp, M. Börner, J. Mohr & C. David, Fabrication of two-dimensional hard X-ray diffraction gratings, Microelectronic Engineering 101 (2013) 12-16.

[19] I. W. Rangelow, Critical tasks in high aspect ratio silicon dry etching for microelectromechanical systems, Journal of Vacuum Science & Technology A 21 (2003) 1550-1562.

[20] L. Romano, M. Kagias, K. Jefimovs & M. Stampanoni, Self-assembly nanostructured gold for high aspect ratio silicon microstructures by metal assisted chemical etching, RSC Advances 6 (2016) 16025-16029.

[21] L. Romano, J. Vila-Comamala, K. Jefimovs & M. Stampanoni, High aspect ratio Si micro-gratings by Metal Assisted Chemical Etching, Journal of Microelectronic Engineering (2016), this issue.

[22] L. D. V. Llona, H. V. Jansen & M. C. Elwenspoek, Seedless electroplating on patterned silicon, Journal of Micromechanics and Microengineering 16 (2006) S1.

[23] J. Ciulik & M. R. Notis, The Au Sn phase diagram, Journal of Alloys and Compounds 191 (1993) 71-78.

[24] H. Schift, Nanoimprint lithography and micro-embossing in LiGA technology: similarities and differences, Microsystem Technologies 20 (2014) 1773-1781.

[25] *NaPa Emerging Nanopatterning Methods, Library of Processes*. (ed H. Schift), 3rd edn, (2014), https://www.psi.ch/lmn/helmut-schift.

[26] L. Battezzati & A. L. Greer, The viscosity of liquid metals and alloys, Acta Metallurgica 37 (1989) 1791-1802.

[27] L. J. Heyderman, H. Schift, C. David, J. Gobrecht & T. Schweizer, Flow behaviour of thin polymer films used for hot embossing lithography, Microelectronic Engineering 54 (2000) 229-245.

[28] N. Yazdani, V. Chawla, E. Edwards, V. Wood, H. G. Park & I. Utke, Modeling and optimization of atomic layer deposition processes on vertically aligned carbon nanotubes, Beilstein Journal of Nanotechnology 5 (2014) 234-244.